\documentclass[12pt]{article}
\begin{document}
\thispagestyle{empty}

%\vspace*{0.2in}
%\begin{flushright}
%\hfill\vbox{\hbox{UA/NPPS-?-04}}
%\end{flushright}
%\vspace{2cm}

%\begin{center}
%\documentstyle[12pt]{article}
%\textheight 9in \textwidth 6.5in \oddsidemargin -.2in \topmargin -.5in
%\renewcommand {\baselinestretch}{1.5}
%\begin{document}
%\begin{center}

\title{{\large{\bf Theoretical evidence for a tachyonic ghost state contribution\\
to the gluon propagator\\ 
in high energy, forward quark-quark `scattering'}}\thanks{To be published in a special volume
of the journal ``Physics of Atomic Nuclei" (``Yadernaya Fizika") honoring the 70th
birthday of Professor Yury A. Simonov.}}
%{\bf WORLD-LINE TECHNIQUES FOR RESUMMING\\
%GLUON RADIATIVE CORRECTIONS}\\
\author {A. I. Karanikas\thanks{email address: akaranik@cc.uoa.gr}$\,$ 
and C. N. Ktorides\thanks{email address: cktorid@cc.uoa.gr 
~~~~~~~~~~~~~~~~~~~~~~~~~~~~~~~~~~~~~~~~~~~~~~~~~~~~~~~~~~~~~~~~~~~~~~~~~~~~~~~~~~~~~~
~~~~~~~~~~~~~~~~~~~~~~~~~~~~~~~~~~~~~~~~~~~~~~~~~~~~~~~~-{\tiny .}~~~~{\bf UA/NPPS-3-04}}\\
\textit{University of Athens, Physics Department Nuclear \& Particle Physics Section}\\
\textit{Panepistimiopolis, Ilisia GR 157--71, Athens, Greece}}
%\end{center}
\maketitle

\vspace{3cm}
%\numberwithin{equation}{section}
%%%%%%%%%%%%%%%%%%%%%%%%%%%%%%%%%%%%%%%%%%%%%%%%%%%%%%%
\begin{abstract}
Implications stemming from the inclusion of non-perturbative, confinining effects, as contained in the 
Stochastic Vacuum Model of Dosch and Simonov, are considered in the context of a, hypothetical,
quark-quark `scattering process' in the Regge kinematical region. In a computation wherein the 
non-perturbative input enters as a correction to established perturbative results, a careful treatment 
of infrared divergencies is shown to imply the presence of an effective propagator associated with the
existence of a linear term in the static potential. An equivalent statement is to say that the
modified gluonic propagator receives contribution from a tachyonic ghost state, an occurence
which is fully consistent with earlier such suggestions made in the context of low energy QCD phenomenology. 
\end{abstract} 
\newpage

{\bf 1. Introduction}

\vspace*{0.2cm}

From the theoretical point of view, forward scattering at very high energies 
(Regge kinematics) presents a situation where one can readily apply eikonal approximation techniques [1]. 
In this context, such processes provide grounds for exploring long distance properties 
of the underlying fundamental theory for the implicated interaction. For the particular case 
of QCD, long distance behavior constitutes a fundamental issue whose exploration is 
not only relevant to high energy processes but also to low energy phenomenology. 

In the present paper, we revisit the (idealized) problem of quark-quark `scattering' in the Regge 
limit, which has been extensively studied within the framework of pQCD [1-5], with the
aim to extend the aformentioned analyses in a direction which takes into account confining aspects of the theory. 
Specifically, we shall rely on the premises of the 
Stochastic Vacuum Model (SVM), pioneered by Dosch and Simonov [6] for the explicit
purpose of accomodating the confinement property of QCD. A similar approach has been pursued by
Nachtmann [7] by employing a different methodology from the one we shall adopt in this work.
Different will also be the the scope of the present analysis. 

The construction of the SVM is motivated by the intention to incorporate established
observations/results regarding the structure of the QCD vacuum [8] into a well defined
theoretical framework.
In particular, it summarizes all that is known and/or surmized about
its properties through a set of three axioms, which are expressed in terms of 
field strength, as opposed to field potential, correlators. The underlying
stochasticity assumption for the vacuum state facilitates the application
of the cumulant expansion [9], which describes the factorization rules for
higher order gluon field strength correlators in terms of two-point ones. 
One of the first results arrived at through the SVM
is the deduction of the area law for the static Wilson loop, i.e. confinement.  
Specific applications of the
SVM scheme, including comparisons with lattice results can be found, e.g., in Ref [10].

A concrete, as well as practical, way to apply the SVM scheme to specific situations has 
been suggested by Simonov [11]. The idea is 
to use the background gauge fixing method [12] and assign the background
gauge fields with the task of becoming the agents of the non-perturbative dynamics. Specifically, 
one employs the gauge potential splitting $A_\mu^a=\alpha_\mu^a\,+\,B_\mu^a$
with the $\alpha_\mu^a$ being associated with the usual perturbative field modes. The 
$B_\mu^a$, on the other hand, enter as {\it dynamical} fields, assigned with the task 
of carrying the non-perturbative physics through field strength correlators which adhere
to the cumulant expansion rules. 

Following our previous work of Ref [5] we find it convenient to employ the FFS-worldline casting 
of QCD, originally pioneered by Fock [13], Feynman [14] and Schwinger [15] and most recently
revived in path integral versions, see Refs [16-18]. The reason for such a choice is that the
eikonal approximation acquires a straightforward realization in this scheme, since,
in the perturbative context at least, it can be readily implemented by restricting one's
considerations to straight worldline paths. As it will turn out, the inclusion of input
from the SVM will produce a kind of deformation of the eikonal paths, 
with low energy consequences, which represent, subleading, perturbative-nonperturbative interference effects.
The conditions which justify the relevant computation will be made explicit in the text.
Suffice it to say, at this point, that it was explicitly demonstrated in Ref [19] that the
FFS-worldline formulation of QCD, in combination with the background gauge field splitting, is ideally suited
for providing a framework within which one can directly and efficiently apply the premises 
of the SVM.   

The main result of this paper is the following: Once (subleading) contributions incorporating nonperturbative, as
induced by the SVM, corrections to the perturbative expressions are taken into account, then a consistent analysis 
of the infrared issues entering the quark-quark high energy forward `scattering' process
reveals that the gluonic propagator 
exhibits a behavior which can be interpreted in terms of the presence of a `tachyonic mass pole'. Arguments
in favor of such an occurence, with important phenomenological
as well as theoretical implications,
have been promoted, from different perspectives, in several papers [20-22]. 

The organization of the paper is as follows. In the next section we present the basic formulas
related to the `amplitude' for the high energy `quark-quark scattering process' in the
forward direction and display their FFS-worldline form. Section 3 focuses
its attention on infrared issues associated with the pertubative-nonperturbative interference effects
under consideration in this study. The resulting expression for the amplitude is shown to be
equivalent to the introduction of an effective propagator, which is associated with the presence of a
linear term in the static potential. Section 4 extends the implications of the aformentioned
result to issues related to renormalization: Modified running coupling constant and summation
of leading logarithms via the Callan-Symanzyk equation. Finally, the technical manipulations 
leading to the main result of section 3 are displayed in the appendix.  

\vspace*{0.5cm}

{\bf 2. Preliminary considerations}

\vspace*{0.2cm}

Consider an idealized quark-quark `scattering' process in the Regge limit, defined 
by $s/m^2\rightarrow\infty$, $s\gg t(=-q^2=\vec{q}_\perp\,^2)$. The amplitude is
given by
\begin{equation}
T_{ii'\,jj'}(s/m^2,q_\perp^2/\lambda^2)=\int d^2b\, e^{-i\vec{q}
\cdot\vec{b}} E_{ii'\,jj'}(s/m^2,1/b^2\lambda^2),
\end{equation}
where $\lambda$ is an `infrared' scale, $b$ is the impact distance in the 
transverse plane to the direction of the colliding quarks (assumed to be travelling along the $x_3$ axis) while
$E_{ii'\,jj'}$, which incorporates the dynamics of the process, 
is specified, in Euclidean space-time, by [5]
\begin{equation}
 E_{ii'\,jj'}\,=\,
    \Biggl\langle {\rm P} \exp \left[
  ig \int_{-\infty}^{\infty} \, d\tau \, v_1 \cdot
  A(v_1\tau)\right]_{ii'}
  {\rm P} \exp \left[
ig \int _{-\infty}^{\infty} \, d\tau \, v_2 \cdot
  A(v_2\tau +b)\right]_{jj'}\Biggr\rangle _A^{\rm conn}\; .
%\label{eq:1}
\end{equation}
The $v_1$ and $v_2$ are constant four velocities characterizing the respective
eikonal lines of the quarks participating in the `scattering process' and `conn' stands
for `connected'. In Minkowski space (and in light cone coordinates) 
one has $v_1\simeq (v_1^+,0,\vec{0}_\perp)$, 
$v_2\simeq (0,v_2^-,\vec{0}_\perp)$,
$b\simeq (0,0,\vec{b}_\perp)$ with $v_1^+\simeq v_2^-\simeq{1\over \sqrt{2}}
\sqrt{s}/ m$.

Employing the gauge field splitting $A_\mu^a=\alpha_\mu^a\,+\,B_\mu^a$, 
the expectation values with respect to field configurations acquire the form 
$<\cdot\cdot\cdot>_A=<\cdot\cdot\cdot>_{\alpha,B}$. Expanding in terms of powers 
of the perturbative field components, one obtains
\begin{eqnarray}
&&E_{ii'\,jj'}=
    \Biggl\langle {\rm P} \exp \left[
  ig \int_{-\infty}^{\infty} \, d\tau \, v_1 \cdot
  B(v_1\tau)\right]_{ii'}
  {\rm P} \exp \left[
ig \int _{-\infty}^{\infty} \, d\tau \, v_2 \cdot
  B(v_2\tau +b)\right]_{jj'}\Biggr\rangle _B^{\rm conn}\;\nonumber\\ &&\quad
-g^2\int_{-\infty}^\infty ds_2 \int_{-\infty}^\infty ds_1
\Biggl\langle {\rm P} \exp \left[
  ig \int_{s_1}^{\infty} \, d\tau \, v_1 \cdot
  B(v_1\tau)\right]_{ik}
  {\rm P} \exp \left[
ig \int _{-\infty}^{s_1} \, d\tau \, v_1 \cdot
  B(v_1\tau)\right]_{li'} \nonumber\\&&\quad\times
{\rm P} \exp \left[
  ig \int_{s_2}^{\infty} \, d\tau \, v_2 \cdot
  B(v_2\tau+b)\right]_{jm}
  {\rm P} \exp \left[
ig \int _{-\infty}^{s_2} \, d\tau \, v_2 \cdot
  B(v_2\tau +b)\right]_{nj'}\nonumber\\&&\quad\times t_{kl}^at_{mn}^b
v_{2\mu}v_{1\nu}iG_{\mu\nu}^{ba}(v_2s_2+b,\,v_1s_1)\Biggr\rangle _B
+{\cal O}(g^4<\alpha^4>).
\end{eqnarray}
Our objective in this paper is to study the behavior of the amplitude as $|b|\rightarrow 0$.
Accordingly, contributions attributed exclusively to the non-perturbative, background terms
will be ignored given that, by definition, they are finite in this limit.
This narrows the expression of computational interest to the following one 
\begin{eqnarray}
E_{ii'\,jj'}&=&-g^2\frac{N_C}{N_C^2-1}t_{ii'}^at_{jj'}^a
\int_{-\infty}^\infty ds_2 \int_{-\infty}^\infty ds_1
v_{2\mu}v_{1\nu}\Biggl\langle {1\over N_C}Tr_AiG_{\mu\nu}(v_2s_2+b,\,v_1s_1)\Biggr\rangle _B
\nonumber\\&&\quad +{\cal O}(g^4<\alpha^4>)+{\cal O}(b^2)
\end{eqnarray}

The propagator $iG_{\mu\nu}^{ba}(v_2s_2+b,\,v_1s_1)\equiv
<\alpha_\mu^b(v_2 s_2+b)\alpha_\nu^a(v_1 s_1)>$ acquires the following worldline expression 
\begin{equation}
iG_{\mu\nu}^{ba}(v_2s_2+b,\,v_1s_1)
=\int_0^\infty dT\int\limits_{\stackrel{x(0)=v_1s_1}{x(T)=v_2s_2+b}}{\cal D}x(t)
e^{-{1\over 4}\int_0^Tdt\dot{x}^2(t)} {\rm P}\left(e^{g\int_0^TdtJ\cdot F 
+ig\int_0^Tdt\dot{x}\cdot B}\right)^{ba}_{\mu\nu}.
\end{equation}
In the above formula $(J\cdot F)_{\mu\nu}= J_{\mu\nu}^{\alpha\beta}F_{\alpha\beta}$, 
with $J_{\mu\nu}^{\alpha\beta}$ the generators 
for the spin-1 representation of the Lorentz group\footnote{Its incorporation into the
worldline form of the propagator serves to signify the spin of the accomodated modes.}. 
It should also be noted that we 
have employed the notation $B_\mu^{ba}=B_\mu^c(t_G^c)^{ab}=-B_\mu^cf^{abc}$.

To close this section let us briefly comment on gauge symmetry related issues. Given the `idealized' 
process under consideration our handling of gauge invariance will be to let the two
worldlines of the `colliding' quarks to extend to infinity in both directions and
impose the boudary conditions $A_\mu[x(t)]\rightarrow 0$ as $t_{{\rm Eucl}}\rightarrow\pm\infty$.
In this way, the overall worldline configuration introduces a Wilson loop in the path
integrals, given that the end-points of the two trajectories can now be 
joined at $+\infty$ and at $-\infty$. These, of course, do not correspond to boundary
conditions for a scattering process {\it per se}, however our only objective in this work
is to extract long distance implications based on the exchanges taking place in the immediate
vicinity of the points of closest approach between the two worldlines. The study of realistic situations
involving the scattering of physically observable particle entities in the Regge limit,
using the presently adopted methodology, is under current consideration and the relevant analysis 
will be presented in the near future.
Finally, concerning the issue of gauge fixing for the $B$ field sector, our choice is prompted by the
intention to rely on field strength correlators for describing nonperturbative dynamics [11].
It, accordingly, becomes convenient to work in the Fock-Schwinger (F-S) gauge [23, 24]. The latter is specified by 

\begin{equation}
B_\mu^a(x)=-\int_{x_0}^xdu_\nu(\partial_\mu u_\rho)F_{\mu\nu}^a(u)
=-(x-x_0)_\nu\int_0^1d\alpha\,\alpha F_{\mu\nu}(x_0+\alpha(x-x_0)).
\end{equation}
Of course, the arbitrary point $x_0$ should not enter any
gauge invariant expression.

\vspace*{0.5cm}

{\bf 3. Infrared issues associated with the propagation of gluonic modes in a confining environment}

\vspace*{0.2cm}

Consider the quantity defined by 
\begin{eqnarray}
&&I(l)\equiv\frac{1}{N_C^2-1} v_{2\mu}v_{1\nu}\langle iTr_AG_{\mu\nu}(l)\rangle_B= 
\frac{N_C}{N_C^2-1} v_{2\mu}v_{1\nu}\nonumber\\&&\times
\int_0^\infty dT\int\limits_{\stackrel{x(0)=0}{x(T)=l}}
{\cal D}x(t)e^{-{1\over 4}\int_0^Tdt\dot{x}^2(t)}\Biggl\langle{1\over N_C}Tr_A{\rm P}
\exp\left(g\int_0^TdtJ\cdot F 
+ig\int_0^Tdt\dot{x}\cdot B\right)_{\mu\nu}\Biggr\rangle_B
\end{eqnarray}
where we have introduced $l_\mu\equiv v_{2\mu}s_2-v_{1\mu}s_1-b_\mu$. 
It describes the propagation of the perturbative gluon modes in the presence of
the background gauge field modes $B_\mu^a$ and, in this sense, it is expected to incorporate
confinement effects associated with the SVM.  

Generally speaking, one would expect that in a study of a physically relevant
process with the full (and proper) inclusion of non-perturbative
effects, no need would arise for the
introduction of an infrared cutoff to regulate the various expressions
entering the computation at long distances. An infrared scale should, in other words,  
naturallly arise suppressing contributions
from the very large distances ($|l|\rightarrow\infty$). 
Given that the object of the present study is to investigate perturbative/non-perturbative 
interference, long distance effects in the (nonphysical) process of quark-quark
high energy `collision' in the forward direction, it becomes necessary to regulate
infrared divergences associated with the upper limit of the $T$-integral.
Our choice of introducing the infrared cutoff is via the relacement
$\int_0^\infty(\cdot\cdot\cdot) dT\rightarrow\int_0^\infty dTe^{-T\lambda^2}
(\cdot\cdot\cdot)$. On a simple dimensional basis and given the
length scales entering the problem, one could make the association $\lambda\propto\sigma|l|$.

Upon expanding the exponential one obtains
\begin{eqnarray}
&&I(l)=\frac{v_1\cdot v_2}{4\pi^2|l|^2}-\frac{2N_C^2}{N^2_C-1}v_{2\mu}v_{1\nu}
\int_0^\infty dT\,e^{-T\lambda^2}\int_0^Tdt_2\int_0^Tdt_1\,\theta(t_2-t_1)\nonumber\\&&\times
\int\limits_{\stackrel{x(0)=0}{x(T)=l}}
{\cal D}x(t)e^{-{1\over 4}\int_0^Tdt\dot{x}^2(t)}\left\{\delta_{\mu\nu}{1\over N_C}Tr_F
\langle g\dot{x}(t_2)\cdot B(x(t_2))g\dot{x}(t_1)\cdot B(x(t_1))\rangle_B\right.
\nonumber\\&&\quad +\left.{2\over N_C}Tr_F\langle gF_{\mu\rho}(x(t_2)
gF_{\rho\nu}^c(x(t_1)\rangle_B\right\}+{\cal O}(\langle g^4F^4\rangle_B)
\end{eqnarray}

Observe, now, that in the F-S gauge, the following relation holds
\begin{eqnarray} 
&&Tr_F\langle gB_{\mu_2}(x(t_2))gB_{\mu_1}(x(t_1))\rangle_B=(x_2-x_0)_{\nu_2}
(x_1-x_0)_{\nu_1}\int_0^1d\alpha_2\alpha_2\int_0^1d\alpha_1\alpha_1
\nonumber\\&&\quad\quad
\times Tr_F\langle gF_{\mu_2\nu_2}(x_0+\alpha_2(x_2-x_0))
gF_{\mu_1\nu_1}(x_0+\alpha_1(x_1-x_0))\rangle_B,
\end{eqnarray}
which brings into play the field strength correlator.

Setting $u_i=x_0+\alpha_ix(t),\,i=1,2$ one writes
\begin{eqnarray}
&&{1\over 2N_C}\langle gF_{\mu_2\nu_2}^c(u_2)gF_{\mu_1\nu_1}^c(u_1)\rangle_B
={1\over N_C}Tr_F\langle\phi(x_0,u_2) gF_{\mu_2\nu_2}(u_2)\nonumber\\&&\quad\times
\phi(u_2,x_0) \phi(x_0,u_1) 
gF_{\mu_1\nu_1}(u_1)\phi(u_1,x_0) \rangle_B\equiv 
\Delta^{(2)}_{\mu_2\nu_2,\mu_1\nu_1}(u_2-u_1),
\end{eqnarray}
where $\phi(x_0,u_i)={\rm P}exp\left(ig\int_{u_i}^{x_0}dv\cdot B(v)\right)$ and is unity in the F-S gauge.
Its insertion serves to underline the gauge invariance of the
field strength correlator.

With the above in place and upon making the redefinition $t_i\rightarrow Tt_i,\,i=1,2$, one determines
\begin{eqnarray}
&&I(l)=\frac{v_1 \cdot v_2}{4\pi^2|l|^2}-\frac{2N_C^2}{N_C^2-1}v_1 \cdot v_2
\int_0^1d\alpha_2\alpha_2\int_0^1d\alpha_1\alpha_1\int_0^1dt_2\int_0^1dt_1\,\theta(t_2-t_1)
\nonumber\\&&\,\,\times\int_0^\infty dTT^2e^{-T\lambda^2}
\int\limits_{\stackrel{x(0)=0}{x(T)=l}}
{\cal D}x(t)e^{-{1\over 4}\int_0^1dt\dot{x}^2(t)}\left\{16\frac{v_{2\mu}v_{1\nu}}{v_2\cdot v_1}
\Delta^{(2)}_{\mu\rho,\rho\nu}(x(t_2)-x(t_1))\right.\nonumber\\&& +
\left.{1\over T^2}\dot{x}_{\mu_2}(t_2)x_{\nu_2}(t_2)
\dot{x}_{\mu_1}(t_1)x_{\nu_1}x_(t_1)
\Delta^{(2)}_{\mu_2\nu_2,\mu_1\nu_1}[\alpha_2x(t_2)-\alpha_1x(t_1)]\right\}+{\cal O}(\langle
g^4F^4\rangle_B).
\end{eqnarray}

On a kinematic basis, the correlator can be represented as follows [6,10]
\begin{eqnarray}
&&\Delta^{(2)}_{\mu_2\nu_2,\mu_1\nu_1}(z_2-z_1)=(\delta_{\mu_2\mu_1}
\delta_{\nu_2\nu_1}-\delta_{\mu_2\nu_1}\delta_{\nu_2\mu_1})
D(z^2)\nonumber\\&&+{1\over 2}\frac{\partial}{\partial z_{\mu_1}}\left[(z_{\mu_2}
\delta_{\nu_2\nu_1}-z_{\nu_2}\delta_{\mu_2\nu_1})D_1(z^2)\right]
+\frac{\partial}{\partial z_{\nu_1}}\left[(z_{\nu_2}
\delta_{\mu_2\mu_1}-z_{\mu_2}\delta_{\nu_2\mu_1})D_1(z^2)\right].
\end{eqnarray}
One notices [10,11] that the first term enters as a distinct feature of the
non-abelian nature of the gauge symmetry (it is not present, e.g., in QED). According
to the premises of the SVM it is associated with the (QCD) string tension $\sigma$ by [10,11]
\begin{equation}
\int_0^\infty dz^2D(z^2)={1\over\pi}\int d^2zD(z^2)\equiv{2\over \pi}\sigma.
\end{equation}
Now, the central objective the present analysis is to determine 
first order contributions to the amplitude coming,
via the SVM, from the non-perurbative/confining sector of QCD. The corresponding, lowest order 
correction is expected, on
dimensional grounds, to be of the string tension $\sigma$. This imlplies, as already pointed 
out by Simonov [25], that we shall set aside the $D_1$ term entering the kinematical analysis
of the correlator, according to Eq (12), which cannot furnish contributing terms of
dimension $[m]^2$. In the appendix the following expression for the quantity $I(l)$ is established
\begin{equation}
I(l)={1\over 4\pi^2}\frac{v_1\cdot v_2}{|l|^2}\left[1+\alpha\sigma|l|^2
\alpha{\rm ln}\left(C{\sigma\over\lambda^2}\right)+{\cal O}(\sigma^2|l|^4)\right],
\end{equation}
where $\alpha\equiv{3\over\pi}\frac{N_C}{N_C^2-1}(1-\kappa)$ with the constant parameter
estimated to be $\kappa\simeq 0.5$. As
pointed out by Simonov [21], the first -and most important- term 
contributing to $\alpha$ comes from the paramegnetic, attractive
interaction between the spin of the gluons with the non-perturbative background field, cf. Eq (5). It
should also be noted that the constant $C$ entering the argument of the logarithm is connected with the
choice of parametrization of $D(z^2)$ see, eg., Ref [7]. In the context of a corresponding result
having to do with an amplitude for a physically relevant, hence
protected from infrared divergencies, process, then any such parameter would disappear.

Suppose the following problem is now posed: Given the above results, which 
pertain to the gluon propagation in a confining environment, 
look for an equivalent, effective particle-like mode propagation, summarizing their
full content. In this spirit, we shall proceed to assess the
possibility that the gist of all we have done up to here can be reproduced via the introduction of an
effective propagator. Following Ref [20] we make the substitution 
\begin{equation}
{1\over k^2}\rightarrow {1\over k^2}+{\mu^2\over k^4}
\end{equation} 
whose additional term signifies the presence of a linear term in the static potential.
Then, one would obtain
\begin{equation}
I(l)=v_1\cdot v_2\int\frac{d^4k}{(2\pi)^4}e^{ik\cdot l}
\left(\frac{1}{k^2}+\frac{\mu^2}{k^4}\right).
\end{equation}
The integral is infrared divergent and should require the introduction of a corresponding
cutoff. Alternatively, one could restrict the validity of the replacement, 
according to Eq (15), to the region $k^2>\mu^2$. Then, one would determine
\begin{eqnarray}
I(l)&=&v_1\cdot v_2\int\limits_{k^2>\mu^2}\frac{d^4k}{(\pi)^2}e^{ik\cdot l}
\left(\frac{1}{k^2}+\frac{\mu^2}{k^4}\right)\nonumber\\&\simeq&
\frac{v_1\cdot v_2}{|l|^2}\frac{1}{(2\pi)^4}\left[1+
\frac{\mu^2|l|^2}{4}{\rm ln}\frac{4}{e\mu^2|l|^2}+{\cal O}(\mu^4|l|^4)\right]\quad {\rm for}\, \mu^2|l|^2<1.
\end{eqnarray}
Comparing the above result with that of Eq (14) one deduces 
\begin{equation}
\mu^2=4\alpha\sigma={3\over\pi}\frac{N_c^2}{N_c^2-1}(1-\kappa)\sigma\simeq 2.15\sigma\simeq0.4\,{\rm GeV}^2,
\end{equation}
in full accord with the phenomenologically determined estimate for the tachyonic `pole' [20-22]. One also
observes that $\lambda\sim|l|\sigma$, as per our original supposition.

Returning to the original, full expression, which provides the full dynamical input
for the amlitude, we write
\begin{eqnarray}
E_{ii'jj'}&\simeq& g^2 t^a_{ii'}t^a_{jj'}v_1\cdot v_2\int_{-\infty}^{+\infty}ds_2
\int_{-\infty}^{+\infty}ds_1{1\over 4\pi^2|l|^2}\left[1+\sigma|l|^2\alpha{\rm ln}
\left(C{\sigma\over \lambda^2}\right)\right]\nonumber\\&\simeq&
-g^2 t^a_{ii'}t^a_{jj'}v_1\cdot v_2\int_{-\infty}^{+\infty}ds_2
\int_{-\infty}^{+\infty}ds_1{1\over 4\pi^2|l|^2}\left[1+\frac{\mu^2|l|^2}{4}{\rm ln}
\left(\frac{4}{e\mu^2|l|^2}\right)\right]\nonumber\\&\simeq& 
-g^2 t^a_{ii'}t^a_{jj'}v_1\cdot v_2\int_{-\infty}^{+\infty}ds_2
\int_{-\infty}^{+\infty}ds_1\int\limits_{k^2>\mu^2}\frac{d^4k}{(2\pi)^2}e^{ik\cdot l}
\left({1\over k^2}+{\mu^2\over k^4}\right).
\end{eqnarray}
Concerning the logarithmic factors entering the above result, it is useful to remark that
the various constants appearing in the arguments are not of any particular importance -at least
in the approximation we have been working- given that they would disappear with the appropriate choice
for the infrared cutoff. More importantly, thinking in terms of the significance
of the above results if they became part of an amplitude corresponding to a physically consistent, hence
protected from infrared divergencies, process, then any dependence from these 
constants should be absent.

Going over to Minkowski space, the previous relation assumes the form
\begin{equation}
E_{ii'jj'}\simeq-g^2 t^a_{ii'}t^a_{jj'}iv_1\cdot v_2\int_{-\infty}^{+\infty}ds_2
\int_{-\infty}^{+\infty}ds_1\int\limits_{k^2>\mu^2}\frac{d^4k}{(2\pi)^2}e^{-ik\cdot l}
\left(-{1\over k^2}+{\mu^2\over k^4}\right).
\end{equation}
Upon observing that 
\begin{equation}
v_1\cdot v_2\int_{-\infty}^{+\infty}ds_2\int_{-\infty}^{+\infty}ds_1
e^{-ik\cdot v_1s_1-ik\cdot v_2s_2}=(2\pi)^2v_1\cdot v_2\delta(k\cdot v_1)\delta(k\cdot v_2)
=(2\pi)^2\coth\gamma\delta(k_+)\delta(k_-),
\end{equation}
where $\gamma$ is determined by
\begin{equation}
\cosh\gamma\equiv \frac{v_1\cdot v_2}{|v_1||v_2|}={s\over 2m^2}\stackrel{ s/m^2\ll 1}{\Rightarrow}
\gamma\simeq\ln(s/m^2)\Rightarrow\coth \gamma\simeq 1.
\end{equation}      
It follows
\begin{equation}
E_{ii'jj'}\simeq-{g^2\over4\pi} t^a_{ii'}t^a_{jj'}i\coth\gamma\, f(b^2\mu^2)
\end{equation}
where
\begin{equation}
f(b^2\mu^2)={1\over\pi}\int\limits_{k_\perp^2>\mu^2}d^2k_\perp e^{ik_\perp\cdot b}
\left({1\over k_\perp^2}+{\mu^2\over k_\perp^4}\right).
\end{equation}

One immediately notices that if $\mu^2b^2\ll 1$, then
\begin{equation}
f(\mu^2b^2)\simeq{\rm ln}\left(\frac{4e}{\mu^2b^2}\right),
\end{equation}
which recovers the known perturbative result -with an infrared cutoff given by
$\lambda^2\equiv\frac{\mu^2}{4e}$.

As b grows, while remaining in the region $\mu^2b^2<1$, one finds
\begin{equation}
f(\mu^2b^2)\simeq {\rm ln}\left(\frac{4e}{\mu^2b^2}\right)\left(1-\frac{\mu^2b^2}{4}\right)+
{1\over2}\mu^2b^2.
\end{equation}
In turn, this gives
\begin{equation}
E_{ii'jj'}\simeq-{g^2\over 4\pi} t^a_{ii'}t^a_{jj'}i\coth\gamma 
\left\{{\rm ln}\left(\frac{4e}{\mu^2b^2}\right)\left(1-\frac{\mu^2b^2}{4}\right)+
{1\over2}\mu^2b^2\right\}.
\end{equation}

\vspace*{0.5cm}

{\bf 4. Summation of large logarithms}

\vspace*{0.2cm}

The presence of terms $\sim g^2{\rm ln}{1/ b^2\mu^2}$, entering through the 
function $f(b^2\mu^2)$, imposes the need of their summation in the perturbative series.
In the absence of the background field, i.e. in the framework of pQCD, it is well
known that such a summation can be accomplished by employing the renormalization
group strategies, which, for the quark-quark `scattering' process under consideration,
can be justified on the basis that $b^{-1}$ plays the role of an ultraviolet cutoff. As Simonov has
shown [11] the presence of the background field does not alter the, relevant for the summation,
Callan-Symanzyk (C-S) equation. The physical basis on which this is so can be articulated by the following
two arguments:
%\begin{itemize}
\begin{enumerate}
%\begin{enumerate}
\item Contributions from the non-perturbative sector do not introduce additional
divergencies, given that they are finite at short distances.
\item Dimension carrying quantities arising from the non-perturbative sector
(correlators) are structured in terms of combinations of renormalization group
invariant quantities $gB$, i.e. they behave as external momenta, as opposed
to masses which are subject to renormalization.
\end{enumerate}
%\end{itemize}
Consequently, the called for renormalization group evolution follows the footsteps of the
procedure employed in the purely perturbative analysis of the same `process' [5]. In this connection
it is recalled [2,26] that the Wilson contour configuration associated with $E_{ii'jj'}$ mixes
with one corresponding to a pair of closed loops resulting from an alternative way of 
identifying the points at infinity [5]. It is associated with 
\begin{eqnarray}
\bar{E}_{ij'ji'} &=&-{\alpha_S\over\pi}c_F(\gamma\coth\gamma-1) \delta_{ij'}\delta_{ji'} f(b^2\mu^2)
\nonumber\\&&+{\alpha_S\over\pi}c_F(\gamma\coth\gamma-1-i\pi\coth\gamma) 
 t^a_{ij'}t^a_{ji'}i\coth\gamma f(b^2\mu^2)+{\cal O}(\alpha_S^2).
\end{eqnarray}
Accordingly, and upon introducing, in shorthand notation, $W_1\equiv\delta_{ii'}\delta_{jj'}+E_{ii'jj'}$ and
$W_2\equiv\delta_{ij'}\delta_{ji'}+\bar{E}_{ij'ji'}$,  the C-S equation assumes the form
\begin{equation}
\left[M\frac{\partial}{\partial M}+\beta(g)\frac{\partial}{\partial g}\right]W_a
=-\Gamma_{ab}W_b,\,\,a,b=1,2,
\end{equation}
with $M$ playing the role of the uv cutoff whose running takes place between some lower scale (at
which corresponding initial conditons are set) and an upper scale set by $1/b$. The anomalous 
dimension matrix $\Gamma_{ab}$, computed in the context of perturbation theory, see Refs [2,3,5], reads
\begin{equation}
(\Gamma_{ab})=\frac{\alpha_s}{\pi}\left(\begin{array}{ll}
       -\frac{i\pi}{N}coth\gamma \, &\, i\pi coth\gamma\\
       -\gamma coth\gamma+1+i\pi coth\gamma\,&\, N(coth\gamma-1)-
\frac{i\pi}{N}coth\gamma \end{array}\right)+{\cal O}(\alpha^2_S).
\end{equation}
What {\it does} change, with respect to the perturbative analysis, on account of the the presence 
of non-perturbative, background contributions is the dependence of the $W_a$ on the $B$-field
correlators, i.e. one has $W_a=W_a[\{\Delta^{(n)}\}, M,g]$. Given that the computation has taken
into account only the two-point correlator, the extra dependence of the $W_a$ will involve the string tension,
cf. Eq (13). The inclusion of this additional dimensional parameter will have its effects on the 
running coupling constant. 

With this in mind, let us recast Eq  (29) in integral form:
\begin{equation}
W_a[\sigma,M_2,g_B(M_2)]=\left\{{\rm Pexp}\left[-\int_{M_1}^{M_2}\frac{dM}{M}\Gamma(g_B(M))\right]
\right\}_{ab}W_b[\sigma,M_1,g_B(M_1)]
\end{equation}
with the path ordering becoming necessary because the anomalous dimension matrices do
not commute with each other. Concerning the integration limits a consistent choice,
given the premises of the present calculation, is to take $M_2=1/b$ and set 
$M_1= 1/b_0$ with $b_0^2\sigma<1$.
It is observed that the non-perturbative input enters the $W_a$ not only through their
explicit dependence on the string constant, but also -which is the most
impotant- through a running coupling constant $g_B$, which obeys the equation
\begin{equation}
M\frac{\partial}{\partial M}g_B(M)=\beta(g_B(M)).
\end{equation}
The solution of the latter calls for initial conditions which are influenced by the presence of the
non-perturbative background and specifically by $\sigma$. Such matters have been studied by Simonov in [25].

Turning our attention to the `deformation' (to the one-loop order) of the running coupling constant, on account
of its additional dependence on the background field, 
we proceed as follows. Knowing the perturbative result to order $\alpha_S$, we go to Eq (31) and present
its solution in the form
\begin{equation}
W_1[\sigma,{1/
b},\alpha_B(1/b)]=\delta_{ii'}\delta_{jj'}-i\coth\gamma\left[\alpha_S(1/b^2_0)
f(b^2_0\mu^2)
\int^{1/b^2}_{1/b_0^2}\frac{d\tau}{\tau}\alpha_S(\tau)\right]
t_{ii'}^a t_{jj'}^a+{\cal O}(\alpha_S^2).
\end{equation}
It follows that
\begin{equation}
\int_{1/b^2_0}^{1/b^2}\frac{d\tau}{\tau}\alpha_B(\tau)=\alpha_S(1/b^2)f(b^2\mu^2)
-\alpha_S(1/b_0^2)f(b^2_0\mu^2)+{\cal O}(\alpha_S^2).
\end{equation}
Upon comparing with Eq (27) one obtains
\begin{equation}
\alpha_B(\tau)=\alpha_S(\tau)\left\{1+{\mu^2\over 4\tau}\left[{\rm ln}\left(\frac{4\tau}{\mu^2}\right )
-2\right]\right\}+{\cal O}(\alpha_S^2).
\end{equation}
It should be noted that
the validity of the above results holds for $\tau/\mu^2>1$ and $\alpha_S(\tau)=\frac{4\pi}{\beta_0}
\frac{1}{\ln(\tau/\Lambda^2)}<1$.
An indicative estimate, on the basis of Eq (35), is that
if $\alpha_S\simeq0.5$, then $\alpha_B\simeq0.5(1+0.05)$. 
Following Refs [3,5], one surmises that the `amplitude' $A$ for the `process' under 
consideration behaves as 
\begin{equation}
A\sim\exp\left[-\frac{N_C}{2\pi}\ln\left(\frac{s}{m^2}\right)
\int_{1/b^2_0}^{1/b^2}\frac{d\tau}{\tau}\alpha_B(\tau)+{\cal O}(\alpha^2_S)\right]\propto
\exp\left[-{\alpha_S\over 2\pi}N_C\ln\left(\frac{s}{m^2}\right)f(b^2\mu^2)\right].
\end{equation}
from which one `reads' a reggeized behavior for the gluon. The difference from the usual, purely
perturbative result is that the function $f(b^2\mu^2)$ is now connected with the modified 
propagator, as per Eq (24).

In conclusion, we have demostrated that the non-perturbative input, through the SVM, to
the analysis of a hypothetical quark-quark `scattering' process in the Regge kinematical region,
produces a result which, in a phenomenological context, has been argued to be extremely attractive
in reproducing low energy hadron phenomenology. In a sense, this investigation could be considered
as a special example, which justifies Simonov's more general argumentation [22] according to which
the perturbative-nonperturbative interference in static QCD interactions at small distances 
imply the presence of a linear term in the potential.

\newpage

\appendix
\setcounter{section}{0}
\addtocounter{section}{1}
\section*{Appendix}
\setcounter{equation}{0}
\renewcommand{\theequation}{\thesection.\arabic{equation}}

Given the set of the defining, worldline formulas given by Eqs (6)-(15) in section 3, we 
proceed to derive derive Eq (16). In the course
of the derivation the various quantities and parameters appearing in the last part of the section are
specified.

Writing
\begin{equation}
D(z^2)=\int_0^\infty dp\tilde{D}(p)e^{-pz^2}
\end{equation}
one determines
\begin{eqnarray}
I(l)&=&{1\over 4\pi^2}\frac{v_1 \cdot v_2}{|l|^2}
\frac{2N_c^2}{Nc^2-1} v_1\cdot v_2\int_0^1d\alpha_2\alpha_2\int_0^1d\alpha_1\alpha_1
\int_0^1dt_2\int_0^1dt_1\,\theta(t_2-t_1)\nonumber\\&&\times
\int_0^\infty dp\,\tilde{D}(p)\,[48Q(p;t_2,t_1)-R(p;t_2,t_1,\alpha_2\alpha_1)]+{\cal O}(<g^4F^4)>_B),
\end{eqnarray}
where we have made the change $t_i\rightarrow Tt_i,\,i=1,2$ and have introduced the quantities
\begin{equation}
Q(p;t_2,t_1)\equiv\left({\pi\over p}\right)^2\int_0^\infty dT\,e^{-T\lambda^2}
 \int\frac{d^4q}{(2\pi)^4}e^{-\frac{q^2}{4p}}
\int\limits_{\stackrel{x(0)=0}{x(1)=l}}
{\cal D}x(t)e^{-{1\over 4T}\int_0^1dt\dot{x}^2(t)}e^{iq\cdot(x(t_2)-x(t_1))}
\end{equation}
and
\begin{eqnarray}
&&R(p;t_2,t_1,\alpha_2,\alpha_1)\equiv\left({\pi\over p}\right)^2\int_0^\infty dT\,e^{-T\lambda^2}
\int\frac{d^4q}{(2\pi)^4}e^{-\frac{q^2}{4p}}(\delta_{\mu_2\nu_2}\delta_{\mu_1\nu_1}
-\delta_{\mu_2\nu_1}\delta_{\mu_1})\nonumber\\&&\times
\int\limits_{\stackrel{x(0)=0}{x(T)=l}}
{\cal D}x(t)e^{-{1\over 4}\int_0^Tdt\dot{x}^2(t)} 
\dot{x}_{\mu_2}(t_2)x_{\nu_2}(t_2)\dot{x}_{\mu_1}(t_1)x_{\nu_1}(t_1)e^{[iq\cdot
(\alpha_2x(t_2)-\alpha_1(t_1)]}
\end{eqnarray}
with $|l|$, as defined in the text.

The above path integrals can be executed by employing standard techniques, given that `particle' action 
functionals are quadradic (plus a linear term) [16-18]. Ignoring terms giving
contributions ${\cal O}(b^2)$ and using condensed notation from hereon, one determines 
\begin{equation}
Q={1\over 16}{1\over p^2}\int_0^\infty dT\,e^{-T\lambda^2}\int\frac{d^4q}{(2\pi)^4}
e^{-\frac{q^2}{4p}-Tq^2G_{12}}[1+{\cal O}(l^2q^2)]
\end{equation}
and
\begin{equation}
R={1\over 16}{1\over p^2}\int_0^\infty dT\,e^{-T\lambda^2}\int\frac{d^4q}{(2\pi)^4}
e^{-\frac{q^2}{4p}-Tq^2K_{12}}[c_0+Tq^2c_1+T^2q^4c_2++{\cal O}(l^2q^2)],
\end{equation}
where the following, one dimensional particle propagator-type quantities have been introduced 
\begin{equation}
\Delta_{12}=\Delta(t_1,t_2) \equiv t_1(1-t_2)\theta(t_2-t_1)+t_2(1-t_1)\theta(t_1-t_2),
\end{equation}

\begin{equation}
G_{12}=G(t_2,t_1)=\Delta(t_2,t_2)+ \Delta(t_1,t_1)- 2\Delta(t_1,t_2)=|t_2-t_1|(1-|t_2-t_1|)
\end{equation}
and
\begin{equation}
K_{12}=K(t_2,t_1)=\alpha_2^2\Delta(t_2,t_2)+\alpha_1^2 \Delta(t_1,t_1)
- 2\alpha_1\alpha_2\Delta(t_1,t_2).
\end{equation} 

The coefficients entering Eq(A.6) are given by the expressions
\begin{equation}
c_0=-72\Delta_{12}\partial_1\Delta_{12}\partial_2\Delta_{12}
\end{equation}
\begin{eqnarray}
c_1&=&48\Delta_{12}\partial_2\Delta_{12}\partial_1K_{12}+24\alpha_1\partial_1\Delta_{12}\partial_2\Delta_{12}
(\alpha_1\Delta_{11}-\alpha_2\Delta_{12})\nonumber\\&&+
24\alpha_2\partial_1\Delta_{12}\partial_2\Delta_{12}(\alpha_2\Delta_{22}-\alpha_1\Delta_{12})
-12(\alpha_1^2\Delta_{12}\partial_1\Delta_{11}\partial_2\Delta_{12}
+\alpha_2^2\Delta_{12}\partial_2\Delta_{22}\partial_1\Delta_{12}\nonumber\\&&
-\alpha_1\alpha_2\Delta_{12}\partial_1\Delta_{11}\partial_2\Delta_{22}
-\alpha_1\alpha_2\Delta_{12}\partial_1\Delta_{12}\partial_2\Delta_{12})\nonumber\\&&
-12(\alpha_2\Delta_{22}-\alpha_1\Delta_{12})(\alpha_1\partial_1\Delta_{11}\partial_2\Delta_{12}
-\alpha_2\partial_1\Delta_{12}\partial_2\Delta_{12})\nonumber\\&&
-12(\alpha_1\Delta_{11}-\alpha_2\Delta_{12})(\alpha_2\partial_2\Delta_{22}\partial_1\Delta_{12}
-\alpha_1\partial_1\Delta_{12}\partial_2\Delta_{12})
\end{eqnarray}
and
\begin{equation}
c_2=24(\alpha_2\Delta_{22}-\alpha_1\Delta_{12})(\alpha_1\Delta_{11}-\alpha_2\Delta_{12})
\partial_1K_{12}\partial_2\Delta_{12}.
\end{equation}

Given the above, the ``paramagnetic" contribution to Eq (A.2) becomes
\begin{eqnarray}
I_p&=&12\frac{2N_c}{Nc^2-1} v_1\cdot v_2\int_0^\infty dp\,\tilde{D}(p)
\int_0^1dt_2\int_0^1dt_1\,\theta(t_2-t_1)Q(p;t_2,t_1)\nonumber\\&&=
{12\over 16}\frac{2N_c^2}{N_c-1}\frac{ v_1\cdot v_2}{4\pi^2}\int_0^\infty {dp\over p}\tilde{D}(p)
\left[\ln\left(4e^{-\gamma_E}{p\over\lambda^2}\right)+{\cal O}(\lambda^2/p)\right]
\end{eqnarray}
and since
\begin{equation}
\int_0^\infty {dp\over p}\tilde{D}(p)=\int_0^\infty dz^2\,\tilde{D}(z^2)
={1\over\pi}\int_0^\infty d^2z\,\tilde{D}(z^2)\equiv{2\over\pi}\sigma,
\end{equation}
Eq (A.9) gives
\begin{equation}
I_p=\frac{3N_c}{N_c^2-1}\frac{ v_1\cdot v_2}{4\pi^2}\sigma\ln\left(C{\sigma\over\lambda^2}\right).
\end{equation}
This furnishes the correction term from the background gauge field contributions entering Eq (14) in the text.
The constant $C$ entering the above result depends on the parametrization of $D(z^2)$. Following the
one of Nachtman [7], one determines $C=39.65$.
The numerical computation of the factor $\kappa$, based on the expressions for the $c_i$, as
given by (A.11),  produces the value $\kappa\simeq 0.5$.

\vspace*{0.5cm}

\begin{center}
{\bf Acknowledgements}
\end{center}\thispagestyle{empty}

\vspace{0.1cm}

The authors are grateful to our colleague, Professor F. K. Diakonos, for programming the
computations for the constants $c_i$. We would also like to acknowledge financial supports through the
research program ``Pythagoras'' (grant 016) and by the General Secretariat of Research and
Technology of the University of Athens.

\end{document}